# Optoelectronic Analysis of Spectrally Selective Nanophotonic Metafilm Cell for Thermophotovoltaic Energy Conversion

Qing Ni, Payam Sabbaghi, and Liping Wang*

School for Engineering of Matter, Transport & Energy,

Arizona State University, Tempe, Arizona, 85287, USA

*Corresponding author's email: liping.wang@asu.edu

**Abstract**

Thermophotovoltaic (TPV) system can directly convert heat into electricity, whose performance is highly dependent on the spectral selectivity of thermal radiation, as only photons with energy above the bandgap can be converted by the photovoltaic cell for power generation. As many studies have focused on wavelength-selective emitters or filters for improving the TPV energy conversion performance, this work theoretically explores a spectrally selective TPV cell based on an asymmetric Fabry-Perot resonance cavity structure with sub-100-nm GaSb layer. The simulated spectral property of the ultrathin nanophotonic cell structure exhibits a high absorption peak above the bandgap due to the interference effect with electromagnetic field enhanced inside the GaSb layer between top and bottom silver electrodes, while the sub-bandgap absorption is as low as a few percent because of high reflectivity of the metal. An absorption enhancement nearly 20 times at particular frequency above bandgap is achieved within the sub-100-nm GaSb layer with the nanophotonic cell structure compared to the free-standing one. Besides, a thin layer of $MoO_x$ is incorporated into the metafilm cell structure as a hole transport layer to consider the charge collection in practice. With rigorous optoelectronic analysis by considering both radiative and nonradiative recombinations (Shockley-Reed-Hall and Auger), the nanophotonic cell is predicted to achieve a TPV efficiency of 22.8% and output power of 0.62 W/cm$^2$ with a black emitter at 1500 K due to spectrally enhanced in-band absorption and low sub-bandgap absorption. With an ideal selective emitter which is hard to achieve in practice along with thermal stability concerns, the efficiency can be further improved to 28% by eliminating sub-bandgap photons. The proposed spectrally selective nanophotonic metafilm cell could be a viable route to achieve high-efficiency and low-cost TPV energy conversion.

## 1. Introduction

Thermophotovoltaic (TPV) technique can directly convert infrared radiation to electricity. Since it can be combined with different kinds of heat sources like concentrated solar energy [1], combustion heat [2], waste heat [3] and so on, TPV technique has a wide range of applications in both military and commercial areas. The key component of a TPV system is the TPV cell, which absorbs the photons emitted from a high-temperature thermal emitter and generates electron-hole pairs to produce electricity. The typical temperature of a thermal emitter is between 1000 and 2000 K, thus low-bandgap III-V semiconductors as well as their compounds, like GaSb, InGaAs, and InGaAsSb which better utilizes longer wavelength radiation, have been widely used as TPV cells. However, due to the mismatch between the thermal radiation spectrum from the emitter and the cell bandgap, the practical TPV systems suffer from low electrical power output and poor conversion efficiency. To enhance the TPV performance, a common way is to use selective emitters based on nanostructures [1, 4-6] which ideally have unity emittance above the bandgap and zero emittance below the bandgap. However, the selective emitters bear the challenges such as strict high-temperature stability usually exceeding 1000 K and sharp cutoff wavelength besides perfect wavelength selectivity. On the other hand, if TPV cells could achieve spectrally selective absorption, i.e., ideally 100% in-bandgap absorptance and zero sub-bandgap absorption, the TPV performance can also be possibly enhanced bypassing some challenges from selective emitters.

Conventional III-V TPV cells are usually several hundred micros in thickness [7], which not only limits their large-scale commercial applications due to high material cost, but also limits their potential to achieve spectral selectivity. In recent years, thin film cells with thickness of several microns or even less have gained increasing attention due to their great potential for low cost due to its low material consumption, ease of manufacturing, reduced weight and increased flexibility [8, 9]. However, due to the much-reduced thickness, the weak light absorption is one of the major challenges of the thin film cells. It has been demonstrated that enhancing the light absorption in thin film cells can be achieved through the use of light trapping structures (textured surfaces, photonic crystals, nanowires, gratings, etc.) based on the excitation of surface plasmons and/or magnetic polaritons [10-15]. Most of these structures are nanostructures with complicated geometries, which prevents low-cost fabrication. Apart from the abovementioned nanostructures, light absorption enhancement can also be realized via planar structures based on anti-reflection effect or cavity resonance [16-18], which are much easier and more affordable to fabricate. Burger et al. [19] proposed a thin film InGaAs TPV cell based on the anti-reflection effect and demonstrated its broadband light absorption enhancement above the bandgap. However, the electrical contact was not considered in their work. To our knowledge, spectrally selective ultrathin (<100 nm) III-V TPV cells based on cavity resonance have not been studied yet.

In conventional TPV cells, p-n junction created by diffusion of dopants at high temperatures is used to separate and extract the photogenerated carriers for generating electricity via the built-in electrical field. However, it might be challenging to form a p-n junction in sub-100-nm cells whose thickness is comparable to the dopant diffuse length. Recently, a novel approach utilizing carrier-selective layers, which allow one type of carrier (electron or hole) passing through while blocking the other type via energy band alignment at the contact regions, has emerged as a cost-effective and efficient alternative to the doped p-n junctions. Although carrier-selective contacts have already been extensively applied in solar cells such as Si, GaAs and InP cells [20-23], this topic of research is only at its infancy for TPV cells. Alcañiz et al. [24] firstly demonstrated a hole selective contact made of transparent substoichiometric molybdenum oxide ($MoO_x$, x<3) on an n-type Ge TPV cell.

In this work, a nanophotonic GaSb cell with sub-100-nm thickness based on a Fabry-Perot cavity structure was proposed and its optoelectronic performance is analyzed for TPV energy conversion with a thin $MoO_x$ layer incorporated as the hole selective layer. The spectral absorptance of the metafilm cell was simulated and the total phase shift was used to elucidate the Fabry-Perot resonance mechanism that leads to spectrally enhanced in-band absorption. Effect of the incident angle on the optical property was also studied. Moreover, with consideration of both radiative and nonradiative recombination mechanisms, the optoelectronic performance of the wavelength-selective nanophotonic cell is evaluated in a simple TPV system with a black or ideal selective emitter at various temperatures. The current-voltage relation, recombination mechanisms, output electric power, net radiative heat flux, and TPV conversion efficiency are discussed in detail.

**2. Nanophotonic design and theoretical background**

2.1 TPV performance evaluation

Figure 1(a) shows the schematic of a simple TPV system with a high-temperature emitter and a cell respectively at temperatures $T_e$ and $T_c$. The emitter gives out photons towards the cell, whereas the incident photons with energy above the bandgap can be absorbed by the cell to excite free electron-hole pairs, thereby generating electricity. In the meantime, thermalization loss exists due to the photons with energy greater than the bandgap as the free carriers excited by these photons dissipate their energy in excess of the bandgap to the phonons, contributing to heat generation. On the other hand, photons with energy below the bandgap cannot excite free carriers but merely cause heating, thus reducing the system efficiency. Therefore, the ideal spectrally selective TPV cell should exhibit narrowband close-to-unity absorptance right above the bandgap with close-to-zero absorptance elsewhere.

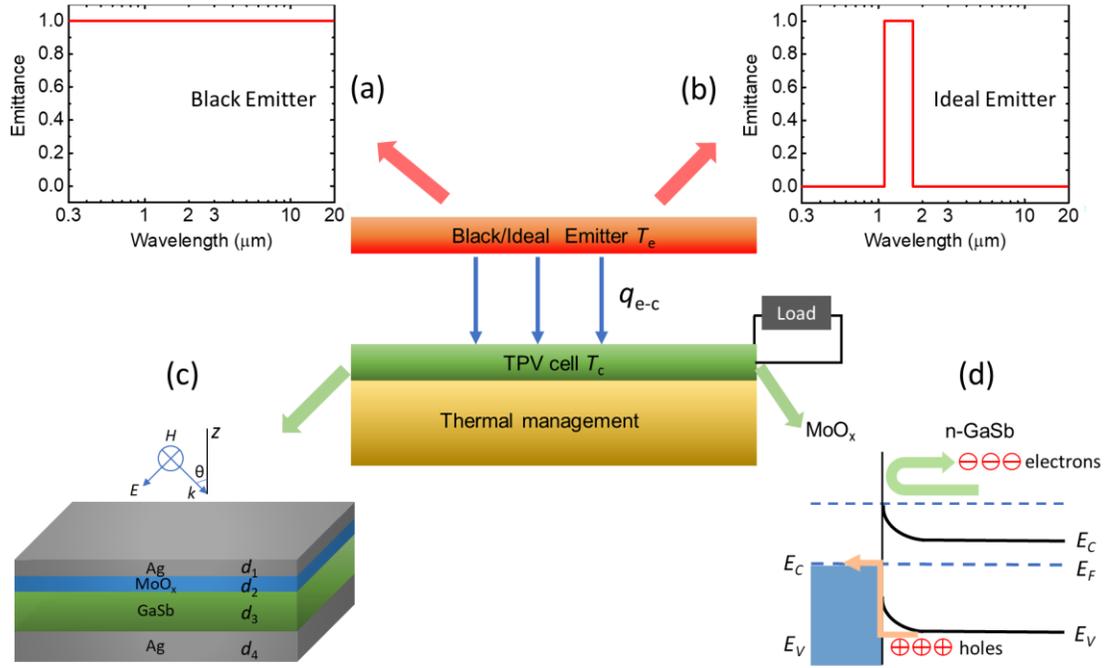

Figure 1. Schematic of a simple TPV system, spectral absorptance of black/ideal emitter, structure of the proposed ultrathin GaSb cell and schematic of the energy band diagram of the MoO$_x$/n-GaSb interface.

To discuss the performance of the proposed ultrathin GaSb cell in a TPV system, the conversion efficiency $\eta$ can be calculated by

$$\eta = \frac{P_e}{q_{in}} \quad (1)$$

where $P_e$ is the maximum output electric power density produced by the cell based on the *J-V* characteristics, and $q_{in}$ is the net radiative heat flux between the emitter and the cell with the same area given

$$q_{in} = \int_0^\infty q_{e\text{-}c,\lambda}\, d\lambda = \int_0^\infty \frac{E_{b\lambda}(T_e) - E_{b\lambda}(T_c)}{\frac{(1-\varepsilon_{e,\lambda})}{\varepsilon_{e,\lambda}} + \frac{1}{F_{e\text{-}c}} + \frac{(1-\varepsilon_{c,\lambda})}{\varepsilon_{c,\lambda}}}\, d\lambda \quad (2)$$

where $q_{e\text{-}c,\lambda}$ is the spectral net radiative heat transfer between the emitter and the cell, and the subscript e(c) represents the emitter (cell). $\varepsilon_{e,\lambda}$ is the spectral emittance of the emitter. A black emitter and an ideal selective emitter are considered here to pair with the proposed nanophotonic

TPV cell. For the black emitter, $\varepsilon_{e,\lambda} = 1$ exists for the whole spectrum (Fig. 1a). For the ideal selective emitter, $\varepsilon_{e,\lambda} = 1$ exists only in a narrow band from $\lambda_1$ to $\lambda_2$ while $\varepsilon_{e,\lambda} = 0$ exists out of this range (Fig. 1b). $\lambda_2$ is fixed at the bandgap of the cell while $\lambda_1$ could be optimized to maximize the efficiency at each emitter temperature. $\varepsilon_{c,\lambda}$ is the spectral emittance of the nanophotonic cell, which equals to the spectral absorptance $\alpha_{c,\lambda}$ according to Kirchhoff's law. $F_{e-c}$ is the view factor between the emitter and the cell, which is assumed to be one in this work for simplicity. $E_{b\lambda}$ represents the blackbody spectral emissive power given by [25]

$$E_{b\lambda}(T) = \frac{2\pi h c_0^2}{\lambda^5 [exp(hc_0/\lambda k_B T) - 1]} \quad (3)$$

where $h$ is the Planck's constant, $c_0$ is the speed of light in vacuum, $k_B$ is the Boltzmann constant and $T$ is the absolute temperature of the blackbody.

2.2 Nanophotonic design of the ultrathin selective metafilm TPV cell

Figure 1(c) depicts the proposed ultrathin GaSb cell. The absorber layer is made of a lightly n-doped GaSb layer with thickness of $d_{GaSb} = 90$ nm. A layer of 20-nm $MoO_x$ thin film is added onto the GaSb layer as a hole selective layer, and a 5-nm Ag film is on top of the $MoO_x$ as the front electrode. The GaSb layer is backed with 200-nm Ag film as the backside electrode, which makes the entire cell structure opaque in the visible and infrared regime. Note that the top Ag film, MoOx/GaSb layer, and the bottom Ag film basically form an asymmetric Fabry-Perot resonance cavity, which could possibly enhance the light absorption inside the GaSb layer with the wave interference effect. The energy band diagram of the $MoO_x$/n-type GaSb interface is shown in Fig. 1(d). Since $MoO_x$ exhibits a high work function (up to 6.6 eV) [26] originating from the oxygen-vacancy-derived defect band inside the cell bandgap, it can serve as a hole transport layer (or electron blocking layer) when placed in contact with n-type GaSb which has a lower work function (4.76 eV) [27] due to the large chemical potential difference [28]. In addition, with a wide bandgap of 3.3 eV [26], $MoO_x$ does not attenuate optical energy that could reduce the light absorption inside the GaSb layer due to its extremely low absorption in the near-infrared.

2.3 Theoretical method for optoelectronic analysis

The spectral directional absorptance of each layer $j$ (1,2,3,4) within the proposed metafilm multilayer structure can be readily calculated via the indirect method detailed in Ref. [29] based on transfer matrix formulation as

$$\alpha'_{\lambda,j} = \frac{S_{jz}(z_{j-1}) - S_{jz}(z_j)}{S_{in,z}} = \frac{1}{\gamma_0} Re \frac{\gamma_j^*}{\mu_j^*} \left\{ \left[ \zeta_1 |C_{j1}|^2 - \zeta_2 |C_{j2}|^2 - 2i\, Im(\zeta_3 C_{j1} C_{j2}^*) \right] \right\} \quad (4)$$

where $S_{jz}$ is the z-component Poynting vectors in the $j$th layer, $S_{in,z}$ is the incident radiant flux in the $z$ direction, $\gamma_0 = 2\pi\lambda \cos\theta_0/c_0$ is the z-component of the free-space wavevector with angle of incidence $\theta_0$, $\zeta_1 = 1 - e^{2\,Re(i\gamma_j)d_j}$, $\zeta_2 = 1 - e^{-2\,Re(i\gamma_j)d_j}$ and $\zeta_3 = 1 - e^{i2\,Im(i\gamma_j)d_j}$, in which $d_j$ is the thickness of the $j$th layer. $C_{j1}$ and $C_{j2}$ are propagation matrix whose expressions can be found in Ref. [29]. The spectral absorptance of the entire metafilm structure is $\alpha'_\lambda = \sum \alpha'_{\lambda,j}$. In the calculation, the dielectric function of Ag is given by a Drude model with parameters from Ref. [30], and the optical constants of $MoO_x$ and GaSb are taken from Refs. [31] and [32], respectively.

The current density ($J$) in the ultrathin n-GaSb cell is a function of the voltage ($V$) and can be obtained by [19]

$$J(V) = J_{ph} - q(R_{rad} + R_{nonrad}) \quad (5)$$

in consideration of photon-generated current $J_{ph}$, radiative recombination rate $R_{rad}$ and nonradiative recombination rate $R_{nonrad}$. The generated current density $J_{ph}$ from the photons radiated by the emitter is calculated by

$$J_{\text{ph}} = \int_0^{\frac{hc_0}{E_g}} \frac{q\lambda}{hc_0} \alpha_{\text{GaSb},\lambda} \varepsilon_{e,\lambda} E_{b\lambda}(T_e) d\lambda \tag{6}$$

where $E_g$ is the bandgap of the cell, $\alpha_{\text{GaSb},\lambda}$ is the spectral absorptance within the GaSb layer, $q$ is the elementary charge. The radiative recombination rate can be calculated by [19]

$$R_{\text{rad}} = exp(\frac{qV}{k_B T_c}) \int_0^{\frac{hc_0}{E_g}} \frac{\lambda}{hc_0} \alpha_{\text{GaSb},\lambda} E_{b\lambda}(T_c) d\lambda \tag{7}$$

The nonradiative recombination rate contributed by Shockley-Reed-Hall (SRH) and Auger recombinations can be calculated by [19]

$$R_{\text{nonrad}} = d_{\text{GaSb}} \frac{n_i}{\tau} exp(\frac{qV}{2k_B T_c}) + d_{\text{GaSb}} (C_n + C_p) n_i^3 exp(\frac{3qV}{2k_B T_c}) \tag{8}$$

where $\tau$ is the SRH lifetime. $n_i = 1.405 \times 10^{12}$ cm$^{-3}$ is the intrinsic carrier concentration of GaSb [33] $C_n = C_p = 5 \times 10^{-30}$ cm$^6$/s are the Auger recombination coefficients of GaSb semiconductor [33]. In this work, we assume an n-type GaSb cell with light doping and an SRH lifetime of $\tau = 10$ ns.

The output power density is calculated from the maximum of the product of the voltage and current by

$$P_e = \max(J \cdot V) \tag{9}$$

## 3. Results and discussion

3.1 Spectrally selective optical absorption above cell bandgap

Figure 2(a) shows the spectral absorptance of the proposed ultrathin nanophotonic GaSb cell with the MoO$_x$ hole-transport layer calculated under normal incidence. Above the GaSb bandgap of 0.72 eV there exhibits a high absorption peak with the maximum absorptance of 0.96 at the incident photon energy of 0.87 eV, while in comparison a free-standing 90-nm GaSb layer could only absorb less than 10% of incident energy, indicating nearly 20 times in absorption enhancement at the incident photon energy of 0.87 eV with the nanophotonic cell structure. With the low sub-bandgap absorptance less than 0.05, the proposed metafilm GaSb cell exhibits excellent spectrally selectivity with almost perfect narrowband absorption above the bandgap, which could efficiently produce photon-excited free carriers for power generation.

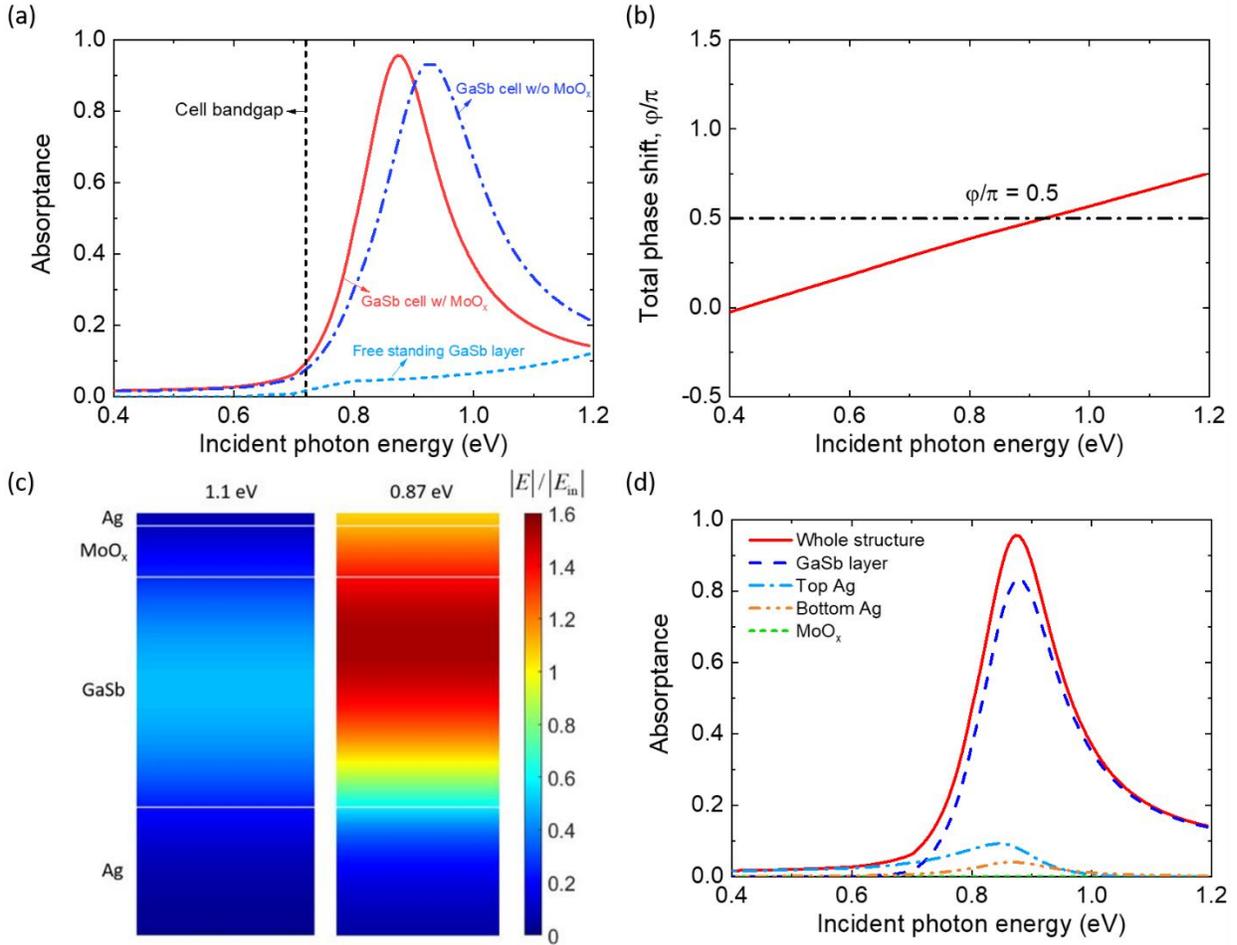

Figure 2. (a) Simulated spectral normal absorptance of the proposed GaSb cell w/ and w/o $MoO_x$. (b) Total phase shift in the cavity of the proposed ultrathin GaSb cell w/o $MoO_x$ layer vs incident photon energy. (c) Normalized magnitude of the electric field at the *x-z* section for the proposed structure. (d) Spectral normal absorptance of each individual layer of the proposed structure.

In order to identify and understand the physical mechanism that is responsible for the spectral absorptance peak above the bandgap from the proposed metafilm GaSb cell, the spectral normal absorptance of a similar structure without the $MoO_x$ layer was calculated with a similar large absorption peak at a slightly higher frequency of 0.93 eV as shown in Fig. 2(a), indicating the wave interference effect from the Ag-GaSb-Ag Fabry-Perot cavity. To verify that, the total phase shift $\psi$ in this metafilm GaSb cell structure without the MoO$_x$ layer is calculated by [34]

$$\psi = (2\beta - \varphi_a + \varphi_s + \varphi_\varsigma)/2 \tag{10}$$

Note that $\varphi_a = arg(r_a)$, $\varphi_s = arg(r_s)$, $\varphi_\varsigma = arg(t_a t_b - r_a r_b)$, where $t_a$ and $r_a$ are respectively the transmission and reflection coefficients from air to the GaSb layer across the top Ag thin film, $t_b$ and $r_b$ are respectively the transmission and reflection coefficients from the GaSb layer to air across the Ag thin film, and $r_s$ is the reflection coefficient from the GaSb to the Ag substrate. $\beta = 2\pi n_3 d_3 \cos\theta_3 /\lambda$ is the phase shift inside the GaSb layer, where $n_3$ is the refractive index of GaSb and $\theta_3$ is the refraction angle inside GaSb. When the total phase shift satisfies $\psi = (m + \frac{1}{2})\pi$ with integer $m$, constructive wave interference occurs inside the GaSb cavity.

Figure 2(b) presents the total phase shift as a function of the incident photon energy at normal incidence for the Ag-GaSb-Ag structure without the MoO$_x$ layer. Clearly the total phase shift $\psi$ equals $\pi/2$ exactly at the incident photon energy of 0.93 eV, directly verifying that the spectral absorption peak at the same frequency stems from the wave interference effect, also known as Fabry-Perot resonance. Note that the interference effect here is different from that for the antireflective coating with thickness as quarter-wavelength on a dielectric substrate because of top absorbing Ag thin film and the GaSb cavity in the proposed metafilm cell structure. When a 20-nm MoO$_x$ layer is added onto the GaSb layer, the total optical thickness of the cavity is increased and the wave interference has to occur with a longer wavelength, which qualitatively explains why the absorption peak frequency redshifts to from 0.93eV to 0.87 eV as observed from the calculated optical absorptance spectra. Figure 2(c) plots the cross-sectional electric field distribution inside the proposed metafilm cell structure with the MoO$_x$ layer at the peak frequency of 0.87 eV and an off-peak frequency of 1.1 eV for comparison. It is clearly observed that the electric field is greatly enhanced within the GaSb and $MoO_x$ dielectric layers as high as 1.6 times of incidence at the peak frequency 0.87 eV, confirming the wave interference effect, while no electric field enhancement is seen at 1.1 eV. Undoubtedly, the Fabry-Perot resonance excited with the proposed metafilm structure achieves the light trapping effect within the sub-100-nm GaSb layer for greatly enhanced photon absorption selectively at frequencies above the bandgap.

Note that only the in-band photons absorbed by the GaSb layer are useful to excite free carriers. Therefore, the absorptance of each layer within the proposed metafilm cell structure is calculated according to Eq. (4) and shown in Fig. 4(d). Clearly most energy absorbed by the whole structure is actually absorbed by the GaSb layer with peak absorptance more than 0.8 at the Fabry-Perot resonance frequency of 0.87 eV, thanks to the low-loss top Ag layer with less than 5% absorptance and the non-absorbing $MoO_x$ layer.

3.2 Polarization and directional independence

To efficiently convert infrared photons from the TPV emitter into electricity, the TPV cell is desired to absorb thermal radiation independently of direction and polarization. However, previous work reported that the multilayered selective emitter made of Au-$SiO_2$-Au asymmetric Fabry-Perot resonance cavity exhibits strong directional dependence with the resonant peak frequency shifts with angle of incidence [16]. To investigate the directional behavior of the proposed ultrathin metafilm GaSb cell with the $MoO_x$ layer, its spectral absorptance was calculated at different incident angles under transverse magnetic (TM) waves and transverse electric (TE) waves as respectively shown in Figs. 3(a) and 3(b). A bright contour band with the close-to-unity absorption due to the Fabry-Perot resonance exists selectively at the same photon energy range (0.77 ~ 0.87 eV) for both polarizations, and it does not change with the incident angles indicating a quasi-diffuse behavior. Note that the absorption band is broadened at large incidence angles only with TM waves because of the Brewster effect [34, 35].

To quantitatively understand the angular-independent behavior of the Fabry-Perot resonance from the metafilm GaSb cell, the total phase shift $\psi$ without the $MoO_x$ layer was analytically calculated based on Eq. (10) as a function of incident photon energy at different incident angles for both polarizations. As clearly shown in Figs. 3(c) and 3(d), the total phase shift does not change with incident angles except for nearly grazing 75° at TM waves, and therefore the resonance condition $\psi = \pi/2$ stays at 0.93 eV frequency regardless of incident angles and polarization states. This unequivocally explains the unique characteristic of direction and polarization independence from the proposed metafilm GaSb cell for efficient photon absorption.

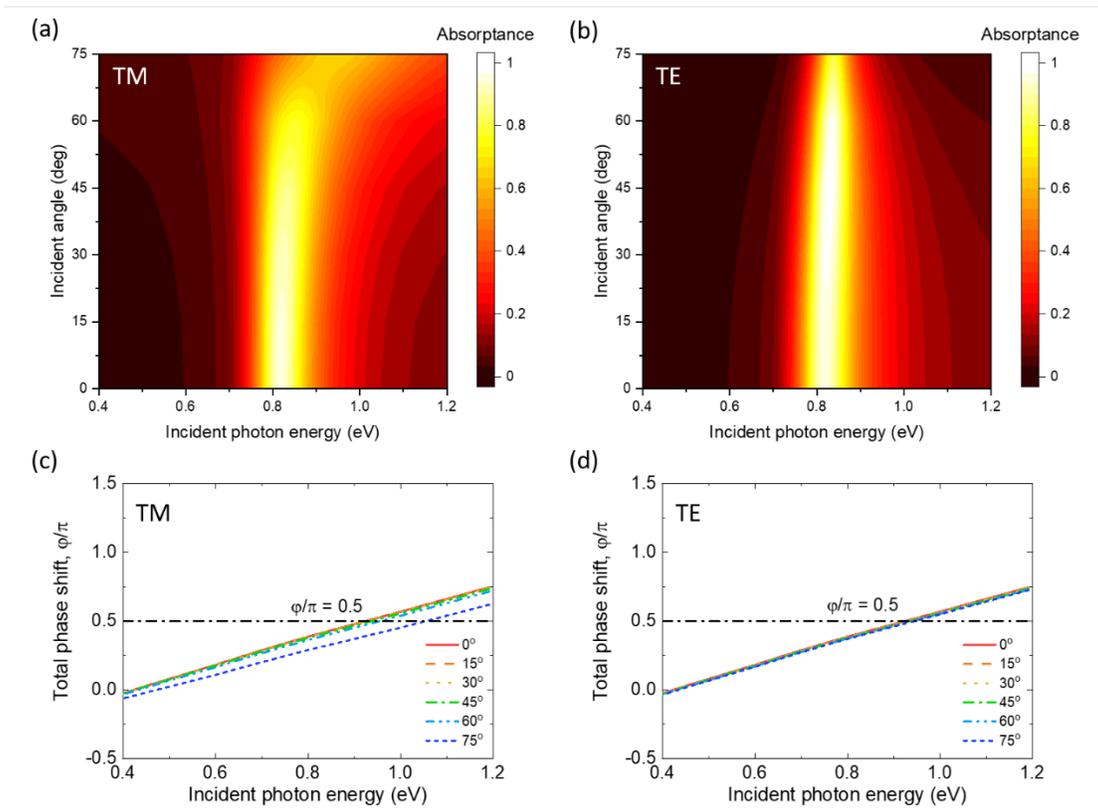

Figure 3. Angular dependence of the spectral absorptance of proposed GaSb cell for (a) TM polarization and (b) TE polarization. Total phase shift at different incident angles for (c) TM polarization and (d) TE polarization.

3.3 Optoelectronic performance for TPV energy conversion

With the enhanced absorption of in-band photons from the proposed ultrathin nanophotonic GaSb cell, the optoelectronic performance for the TPV energy conversion is analyzed by pairing the metafilm cell with a black or ideal emitter at different temperatures in a simple TPV system as depicted in Figure 1. The radiative and nonradiative recombination rates of the proposed metafilm GaSb cell are first studied as a function of cell voltage at fixed cell temperature $T_c = 300$ K. As seen in Fig. 4(a) the nonradiative recombination due to SRH and Auger mechanisms is dominant when the voltage is less than 0.33 V, while the radiative recombination from photoluminescence effect prevails at larger voltage exceeding 0.33 V. The observed recombination behavior here agrees well with Zn-diffusion fabricated GaSb bulk cell [36] in which the nonradiative recombination is dominant.

Figure 4(b) presents the current-voltage (J-V) characteristic from the metafilm GaSb cell paired with a black emitter at $T_e = 1500$ K. Without considering any recombination, the current density is a constant value of 2.14 A/cm$^2$ regardless of the cell voltage from Eq. (6). When both radiative and nonradiative recombinations are taken into account, the current density starts to drop at the cell voltage around 0.2 V, and reaches zero at the open-circuit voltage of 0.39 V. Interestingly, if only radiative recombination is factored in, the J-V curve is little affected with a slightly higher open-circuit voltage of 0.41 V, indicating that the radiative recombination dominates the open-circuit voltage. Indeed, the photoluminescence quantum yield (PLQY), which is the ratio of the radiative recombination to the total (radiative and nonradiative) recombination, is about 80% at cell voltage of 0.4 V as shown in Figure 4 (a). A larger PLQY value with stronger radiative recombination would increase the open-circuit voltage and the power generation [37].

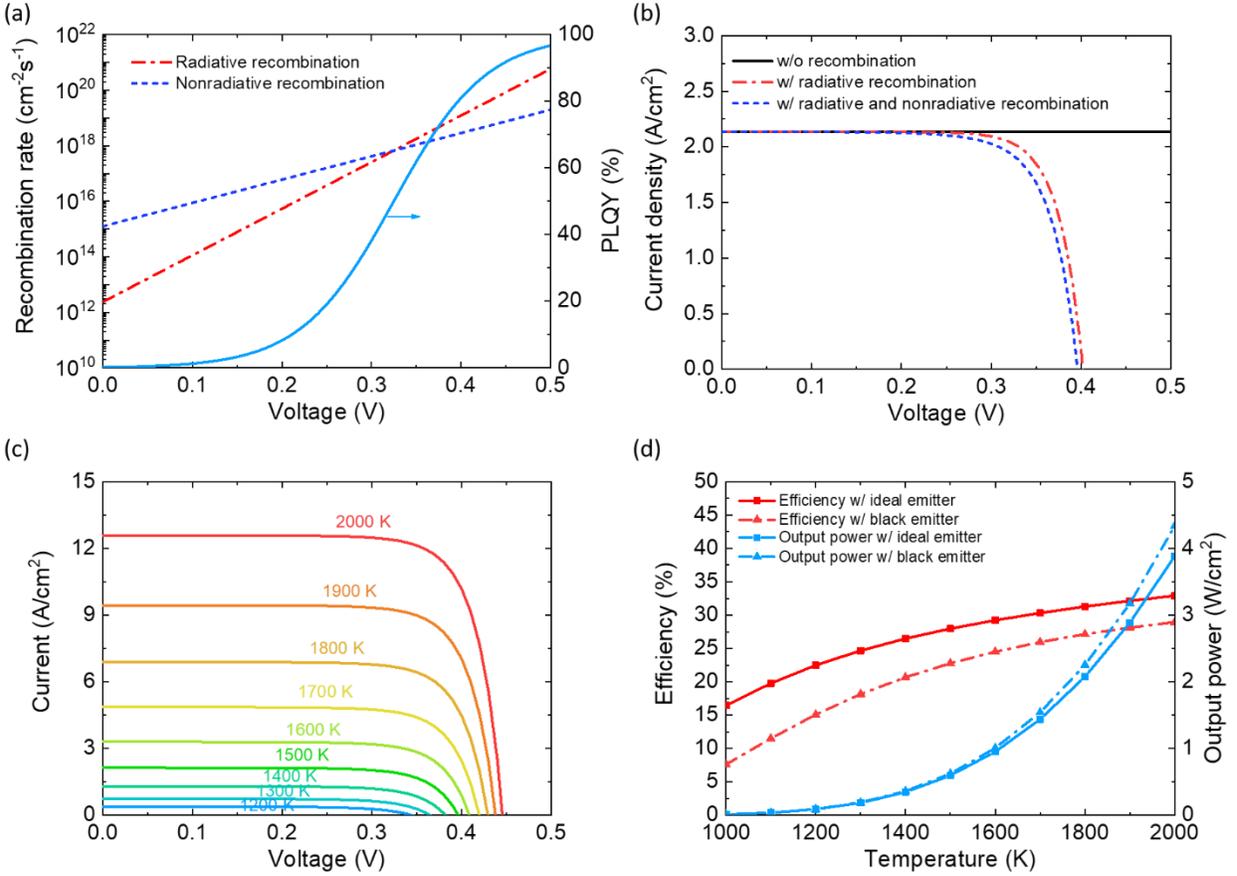

Figure 4. (a) Recombination rates of radiative, and nonradiative recombination. (b) Current density w/ and w/o considering recombination. (c) Simulated J-V curves of the proposed nanophotonic GaSb cell at different black emitter temperatures. (d) Efficiency and output power at different temperatures from 1000 K to 2000 K with either a black emitter or an ideal selective emitter.

Figure 4(c) displays the photon-generation current density of the metafilm GaSb cell with the black emitter at different temperatures with both radiative and nonradiative recombinations considered. As the emitter temperature rises from 1200 K to 2000 K, the short-circuit current density ($J_{sc}$) escalates from 0.4 to 12.6 A/cm$^2$ while the open-circuit voltage ($V_{oc}$) slightly increases from 0.36 to 0.45 V. This is simply because more photons from the emitter become available above the cell bandgap as the blackbody spectrum shifts to higher frequencies with increasing temperature, generating larger current density $J_{ph}$. Note that both radiative and nonradiative recombination rates are independent of emitter temperature.

To evaluate the TPV performance, Figure 4(d) predicts the conversion efficiency for the proposed ultrathin nanophotonic GaSb metafilm cell paired with a black or ideal emitter at various temperatures. With the black emitter at temperatures from 1000 K to 2000 K, the ultrathin metafilm cell converts heat to power with efficiency from 7.6% to 29%, thanks to the high spectral selectivity of the metafilm cell with very low sub-bandgap absorption as shown in Fig. 2(a). With the ideal emitter whose short-cutoff wavelength $\lambda_1$ was optimized at each temperature, the TPV efficiency from the same nanophotonic metafilm cell could be further improved by another absolute 4% to 9%, reaching 33% efficiency at 2000 K. In particular, at the median temperature of 1500 K, the proposed GaSb metafilm cell could achieve a TPV efficiency of 22.8% and an output power of 0.62 W/cm$^2$ with the black emitter, or efficiency of 28% and output power of 0.59 W/cm$^2$ with the ideal selective emitter. The slightly better performance with the ideal emitter than the black one could be understood by the elimination of sub-bandgap photons and thereby reduction of net radiative heat flux with almost the same power output, while the black emitter yields a slightly higher power output particularly at high temperatures due to its broader in-band emission than the narrowband ideal emitter. Note that the ideal selective emitter is very difficult to achieve in actual sample fabrication, and most of the reported selective emitters were made of nanostructures with complex geometries from complicated and high-cost fabrication preprocess. Moreover, the thermal stability of the selective emitters at high temperatures over 1500 K for long time operation is another challenge yet to overcome. The proposed nanophotonic metafilm cell with high spectral selectivity without high-temperature concerns could be a viable route to achieve high-efficiency and low-cost TPV energy conversion.

## 4. Conclusion

In summary, we have theoretically demonstrated a spectrally selective ultrathin GaSb cell based on a nanophotonic structure for TPV energy conversion. By utilizing the Fabry-Perot resonance, strong light absorption close to 90% above the bandgap is achieved inside the sub-100-nm GaSb layer, elucidated by the total phase shift, while the sub-bandgap absorption is suppressed by highly reflective metal electrodes without resonance. A thin $MoO_x$ film was considered into the optical modeling as a hole-selective layer for practical charge collection. Rigorous optoelectronic analysis predicted a TPV efficiency of 22.8% and an output power of 0.62 W/cm$^2$ using the proposed ultrathin GaSb cell even paired with a 1500-K black emitter, with considering both the radiative recombination and nonradiative recombination. This work will pave the way for the spectrally selective ultrathin TPV cell and will promote the development of the low-cost and high-efficiency TPV devices.

**Acknowledgements**

This work was supported in part by National Science Foundation (Grant No. CBET-1454698) and Air Force Office of Scientific Research (Grant No. FA9550-17-1-0080). P.S. would like to thank ASU Graduate College for providing the PhD Completion Fellowship.

**Data Availability**

The data that support the findings of this study is available from the corresponding author upon reasonable request.